# Analytical One-Dimensional Model of Drop Ejection from a Micro-Size Nozzle

Leonid Pekker

FujiFilm Dimatix, Inc. Lebanon, New Hampshire 03766, USA, leonid.pekker@fujifilm.com

**Abstract**

In this article, we construct a novel one-dimensional model of drop ejection from a micro-size nozzle due to a short pressure pulse applied to the liquid in the nozzle. The pressure pulse supplies the kinetic energy to the perturbed liquid-bulge squeezed from the nozzle, which then ballistically lengthens forming a ligament. The Plateau-Rayleigh instability forms a neck in the ligament at the nozzle, leading to detachment of the ligament from the nozzle, which then collapses in a drop. This drop formation sequence is typical for drop-on-demand printheads in which the drop is ejected from the nozzle by a short pressure pulse at the needed moment of time when it should reach the substrate. The model calculates the velocity of the droplet, length of the ligament vs. time, and the time when the ligament detaches from the nozzle as a function of the exit radius of the nozzle, the volume of the droplet, the time that the volume of the droplet is squeezed from the nozzle, viscosity, surface tension, and mass density of the liquid drop. The model also calculates a criterion for drop ejection from the nozzle.

Key Words: Drop formation, inkjet printhead, drop-on-demand printhead

## I. Introduction

Drop formation is a key element in the inkjet printhead technology. The CFD, computation fluid dynamics, simulation is the most accurate tool for modeling drop formation in printheads [1]. However, these simulations are very intensive and demand significant computation time because small time and spatial scales are needed to model the motion of liquid inside the nozzle, the formation of the ligament, the detachment of the ligament from the nozzle, and the collapsing of the ligament into droplets. Therefore, it is beneficial to use simple analytical models of drop formation as a tool for down selecting the design parameters for CFD modeling. This can dramatically decrease the needed CFD simulation time to find an optimal design for drop-on-demand inkjet printheads. Using such analytical models would improve our understanding of physical processes of drop formations in general as well.

In drop-on-demand inkjet printheads, the drop is ejected from the nozzle by applying a short pressure pulse which enables a droplet to be ejected from the nozzle and reach the substrate. When pressure pulses are not present inside the jet, the liquid is held inside of the nozzle by surface tension. In typical drop-on-demand printheads, the fluid is ejected from the nozzle during a very short period of time when the pressure pulse is applied to the liquid in the nozzle. The pressure spike supplies the kinetic energy to the liquid bulge squeezed from the nozzle, which then ballistically lengthens forming a long cylindrical ligament. Since the volume of the ligament does not change significantly after the pressure pulse is complete, the velocity of the liquid in the ligament at the exit of the nozzle is much smaller compared to the velocity of the liquid at the tip of ligament. Therefore, in the ballistic regime, the cross-sectional radius of the ligament decreases with time. Because of the Plateau-Rayleigh instability, at some point during the lengthening of the ligament, a neck forms in the ligament at the exit of the nozzle exit plane, with the ligament continuing to decrease in diameter, leading to detachment of the ligament from the nozzle exit plane. The detached ligament then collapses into a drop. The time of the ballistic lengthening of the ligament and the time of development of the Plateau-Rayleigh instability in such printheads are usually much larger than the time of pressure pulse. This allows us to separate the process of squeezing liquid from the nozzle from the processes of the ballistic lengthening of the ligament and the development the Plateau-Rayleigh instability.

In this article, we suggested an analytical axisymmetric model of drop ejection from a micro-size nozzle in which we take advantages of different time scales in the process of drop ejection in the drop-on-demand printheads. Input parameters of the model are: $R_{nozzle}$ – the exit radius of the nozzle, $V_{drop}$ – the volume of the ejected drop, $\tau_{squeeze}$ – the time that the volume of the droplet, $V_{drop}$, is squeezed from the nozzle ($\tau_{squeeze}$ is similar to the duration of pressure pulse, $\tau_{pulse}$), and $\rho$, $\gamma$, $\mu$ – the mass density, the surface tension, and the viscosity of the liquid. Output parameters are: $\tau_{break-off}$ – the time of detachment the ligament from the nozzle, $L_{break-off}$ – the length of the ligament when the detachment occurs, and $v_{drop}$ – the velocity of the drop. The model is presented in Section 2; numerical results and comparison with experiment in Section 3, and conclusions are given in Section 4.

## II. Model of Drop Ejecting from a Micro-Size Nozzle

The structure of this Section as follows: In Section A, we present the assumptions of the model; in Section B, the ballistic regime describing the lengthening of the ligament; in Section C, we the Platea-Rayleigh instability describing the process of breaking-off the ligament from the nozzle; and, in Section D we present the model and the algorithm of the ligament break-off.

### A. Assumptions of the model

Schematics of our axisymmetric analytical model of drop ejection from a micro-size nozzle is shown in Fig. 1. Panel a shows the liquid bulge squeezed from the nozzle due to the pressure pulse applied to the liquid in the nozzle. In the model, we assume: (1) all volume of the drop is squeezed from the nozzle during the $\tau_{squeeze}$ time, and that after that time, $t \geq \tau_{squeeze}$, no liquid exchange occurs between liquid in the nozzle and liquid in the bulge. It means that the velocity of the liquid at the exit of the nozzle is zero for $t \geq \tau_{squeeze}$; (2) the bulge is a cylinder with the radius equal to the exit radius of the nozzle, $R = R_{nozzle}$, and that the bulge meniscus is "flat". Thus, the height of the "cylindrical" bulge at $t = \tau_{squeeze}$ is

$$L_0 = \frac{V_{drop}}{\pi R_{nozzle}^2}. \tag{1}$$

Kinetic energy transferred from the pressure pulse causes the bulge to ballistically lengthen as is illustrated in Panel b. At this stage, we further call the lengthened bulge a ligament. As in Panel a, we also assume that (3) the ligament is cylindrical all the time. Since the volume of the ligament is fixed (it is equal to $V_{drop}$), with time, the radius of the ligament decreases, the ligament surface area increases, and the ligament decelerates because the initial kinetic energy of the bulge is spent on work against the surface tension forces and the viscosity forces. Because of the Plateau-Rayleigh instability, at some point during the lengthening of the ligament, a neck in the ligament starts to form at the exit of the nozzle (see Panel c) which decreases in diameter to zero, leading to detachment of the ligament from the nozzle (see Panel d), and then the detached ligament collapses into a drop (Panel e). However, if the initial kinetic energy of the bulge is not large enough, the ligament can rebound back to the nozzle before the Plateau-Raileigh instability is fully developed and, therefore, the ligament may not detach from the nozzle. We also made the following assumptions: (4) the height of the bulge, Eq. (1), is much shorter compared to the maximal length of the ligament, $L_0 \ll L_{break-off}$; (5) $L_{break-off} \gg R_{nozzle}$, Fig. 1; (6) the longitudinal velocity of the liquid in the bulge, and later in the ligament, is uniform across the jet. Applying the liquid volume conservation law to cylindrical jet we obtain, Panel b in Fig.1,

$$\pi R^2 v_z(z) \Delta t = 2 \pi R z \frac{dR}{dt} \Delta t \rightarrow v_z(z) = \frac{2 z}{R} \frac{dR}{dt} \rightarrow$$

$$\rightarrow v_z(z) = v_{face} \frac{z}{L} \quad \text{and} \quad \frac{dR}{dt} = v_{face} \frac{R}{2L} \tag{2}$$

where $v_{face}$ is the face velocity of the bulge / ligament, Fig. 1, and $L$ is length of the bulge / ligament; equation for $v_{face}$ is trivial:

$$v_{face}(t) = \frac{dL}{dt}. \tag{3}$$

To describing the dynamic of the ligament, we need to know $v_0$, the face velocity of the bulge at $t = \tau_{squeeze}$; in the model, we take it as:

$$v_0 = \frac{4L_0}{\tau_{squeeze}}, \tag{4}$$

assumption (7). In the model, we also assume that the dynamics of the lengthening the ligament in the Plateau-Rayleigh instability regime is the same as in the ballistic regime, assumption (8). In other words, the development of the neck, Fig. 1 Panel c does not affect much the lengthening the ligament. In the model, we neglect the air drag, assumption (9).

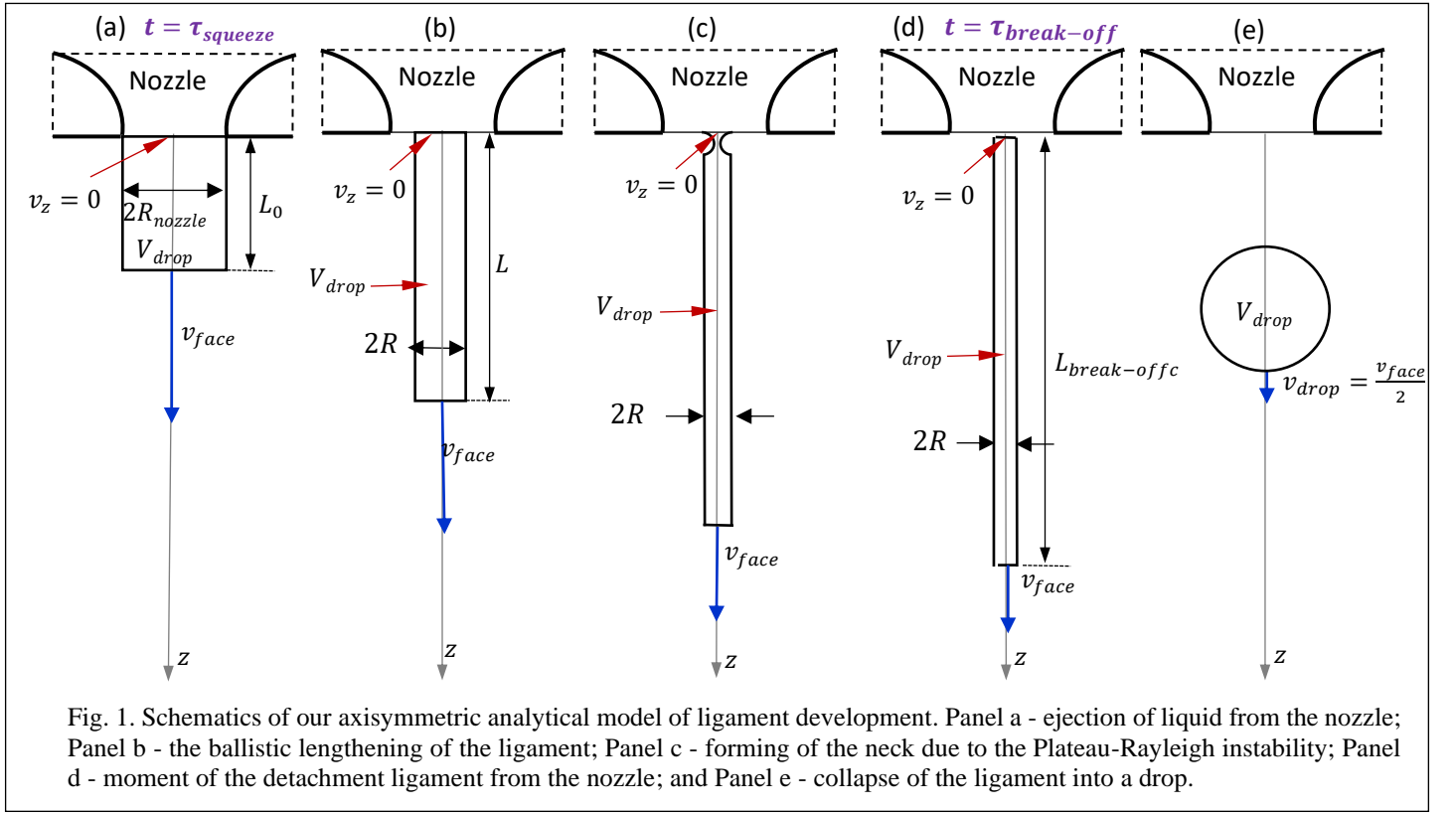

Fig. 1. Schematics of our axisymmetric analytical model of ligament development. Panel a - ejection of liquid from the nozzle; Panel b - the ballistic lengthening of the ligament; Panel c - forming of the neck due to the Plateau-Rayleigh instability; Panel d - moment of the detachment ligament from the nozzle; and Panel e - collapse of the ligament into a drop.

**B. Ballistic regime of lengthening of the ligament**

As follows from Eq. (2), when $R \ll L$, in the ballistic expansion of the ligament regime, the radial velocity of the liquid in the ligament is much smaller compared to the liquid longitudinal velocity, which allows us to utilize a slender jet approximation to describe the dynamics of the ligament. Using the standard slender jet equations [2, 3], we derive in the Appendix the following ordinary differential equation describing the length of the ligament as a function of time:

$$\frac{d}{dt}\left(\frac{\rho V_{drop}}{6}\left(\frac{dL}{dt}\right)^2 + 2\gamma\left(\pi V_{drop}\right)^{1/2}\sqrt{L}\right) = -\frac{3\mu V_{drop}}{L^2}\left(\frac{dL}{dt}\right)^2. \tag{5}$$

In the RHS of Eq. (5), the first term in the parentheses is the kinetic energy of the ligament and the second one is the ligament surface tension energy; the RHS of Eq. (5) describes the kinetic energy plus the surface tension energy loss rate due to the viscosity forces. The boundary conditions for this equation are:

$$(L)_{t=\tau_{squeeze}} = L_0, \tag{6}$$

$$\left(\frac{dL}{dt}\right)_{t=\tau_{squeeze}} = v_0, \tag{7}$$

which correspond to the length and the face velocity of the ligament at $t = \tau_{squeeze}$, Eqs. (1) and (4).

Let us reduce Eq. (5) to the following view:

$$\frac{d}{dt}\left(\frac{\rho V_{drop}}{6}\left(\frac{dL}{dt}\right)^2 + 2\gamma(\pi V_{drop})^{1/2}\sqrt{L}\right) = -\frac{3\mu V_{drop}}{L^2}\left(\frac{dL}{dt}\right)^2 \rightarrow$$

$$\rightarrow \frac{\rho V_{drop}}{3}\frac{dL}{dt}\frac{d^2L}{dt^2} + \sqrt{\pi V_{drop}}\gamma\frac{1}{\sqrt{L}}\frac{dL}{dt} = -\frac{3\mu V_{drop}}{L^2}\left(\frac{dL}{dt}\right)^2 \rightarrow$$

$$\rightarrow \frac{d^2L}{dt^2} + \left(\frac{3\sqrt{\pi}\gamma}{\rho\sqrt{V_{drop}}}\right)\frac{1}{\sqrt{L}} + \left(\frac{9\mu}{\rho}\frac{1}{L^2}\right)\frac{dL}{dt} = 0. \tag{8}$$

As one can see, Eq. (8) is an equation of a non-linear oscillator with dissipation. Thus, the dynamics of the ligament is described by Eq. (8) along with BC (6) and (7).

Fig. 2 presents a solution of Eq. (8) for $V_{drop} = 90 \cdot 10^{-15}\ m^{-3}$, $\rho = 10^3 kg/m^3$, $\gamma = 0.073\ N/m$, $\mu = 0.001\ Pa\ sec$, $\tau_{squeeze} = 10\mu s$, and $R_{nozzle} = 25\mu m$, which are typical for drop-on-demand printheads. As one can see from Fig. 2, at time of $0.000695\ sec$, the ligament stops and rebounds back to the nozzle. The calculated radius of the ligament and the ligament length at this moment of time are, correspondingly, $2.403 \cdot 10^{-6}$m and $4.958 \cdot 10^{-3}$m. This result does not reflect physical reality, but since the ligament is not stable because of the Plateau-Rayleigh instability, it can break-off from the nozzle before rebounding back to the nozzle, as in Fig. 1, panels c and d.

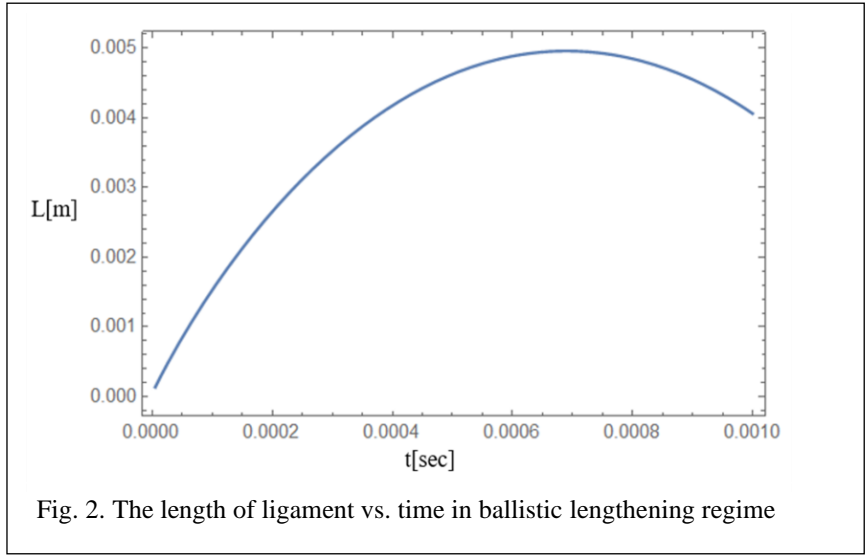

Fig. 2. The length of ligament vs. time in ballistic lengthening regime

## C. Plateau-Rayleigh instability

In [4], Rayleigh obtained the dispersion equation for the Plateau-Rayleigh instability in the case of an infinite cylindrical liquid column perturbed by a small sinusoidal wave, Fig. 3:

$$R(z,t) = R_0 + \delta R_0 \exp(i\omega t - ikz) \quad \text{where } \delta R_0 \ll R_0 \tag{9}$$

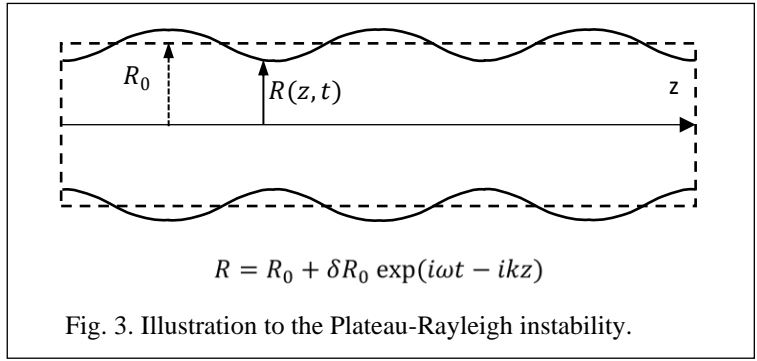

$$R = R_0 + \delta R_0 \exp(i\omega t - ikz)$$

Fig. 3. Illustration to the Plateau-Rayleigh instability.

In [5], the author presents the Rayleigh's dispersion equation in compact form:

$$(1-\tilde{k}^2)\cdot(i\tilde{k})\frac{J_1(i\tilde{k})}{J_0(i\tilde{k})} = \left(1-2i\frac{\tilde{\mu}\cdot\tilde{k}^2}{\tilde{\omega}}\right)\tilde{\omega}^2 - i\tilde{\mu}\tilde{\omega}\tilde{k}^2\left(1-2i\frac{\tilde{\mu}\cdot\tilde{k}^2}{\tilde{\omega}}\right)\cdot$$

$$\left(\left(1-\frac{J_2(i\tilde{k})}{J_0(i\tilde{k})}\right) - \frac{\left(1+i\frac{\tilde{\omega}}{\tilde{\mu}\cdot\tilde{k}^2}\right)^{1/2}}{\left(1+i\frac{\tilde{\omega}}{2\tilde{\mu}\cdot\tilde{k}^2}\right)} \cdot \frac{J_1(i\tilde{k})}{J_1\left(i\tilde{k}\left(1+i\frac{\tilde{\omega}}{\tilde{\mu}\cdot\tilde{k}^2}\right)^{1/2}\right)} \left(\frac{J_0\left(i\tilde{k}\left(1+i\frac{\tilde{\omega}}{\tilde{\mu}\cdot\tilde{k}^2}\right)^{1/2}\right) - J_2\left(i\tilde{k}\left(1+i\frac{\tilde{\omega}}{\tilde{\mu}\cdot\tilde{k}^2}\right)^{1/2}\right)}{J_0(i\tilde{k})}\right)\right), \tag{10}$$

$$\tilde{k} = kR_0 = \frac{2\pi R_0}{\lambda}, \quad \tilde{\mu} = \frac{\mu}{\sqrt{\gamma \cdot \rho \cdot R_0}}, \quad \tilde{\omega} = \omega \sqrt{\frac{\rho \cdot R_0^3}{\gamma}}. \qquad (11)$$

Here $\tilde{\omega}$ is the dimensionless frequency; $\tilde{k}$ is the dimensionless wave number; $\tilde{\mu}$ is the Plateau-Rayleigh's dimensionless viscosity parameter; $J_0$, $J_1$, and $J_2$ are the Bessel functions of the zero, first, and second orders, respectively. As follows from Eq. (9), if $\text{Im}(\tilde{\omega}) < 0$ the wave is unstable – its amplitude increases with time; and if $\text{Im}(\tilde{\omega}) > 0$ the wave is damping – with time its amplitude decreases to zero.

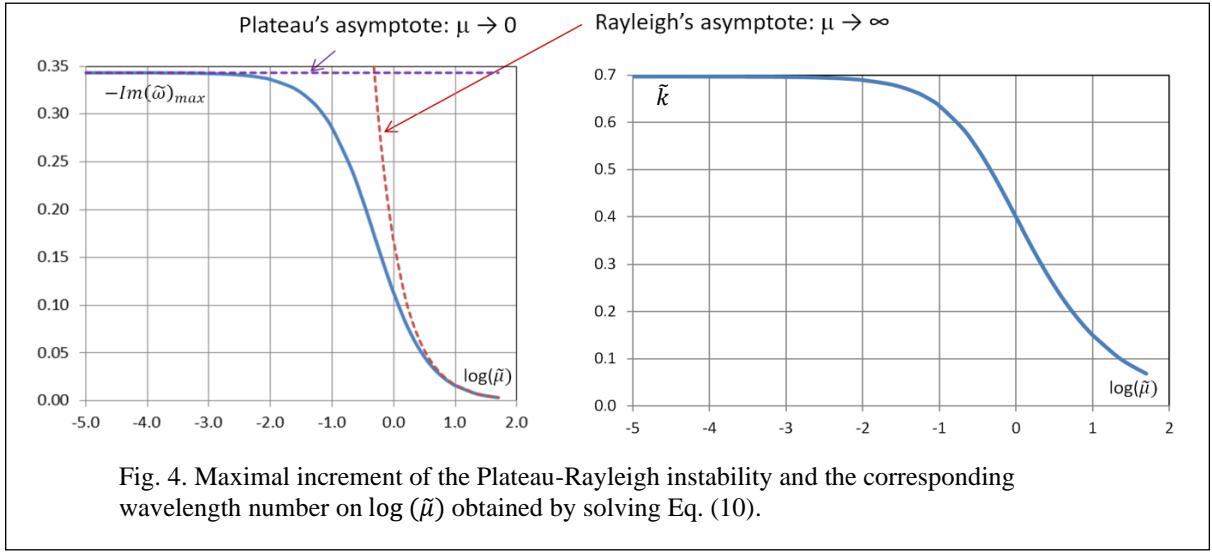

Fig. 4. Maximal increment of the Plateau-Rayleigh instability and the corresponding wavelength number on log ($\tilde{\mu}$) obtained by solving Eq. (10).

In [5], the author calculates the maximum increment of the Plateau-Rayleigh instability and the corresponding wavenumber at given $\tilde{\mu}$; the results are presented in Fig. 4. As expected, with an increase in dimensionless viscosity parameter $\tilde{\mu}$, the increment of the instability $-\text{Im}(\tilde{\omega})_{max}$ and the corresponding wavenumber $\tilde{k}(\tilde{\omega})$ decrease. For our model, it is convenient to consider the Plateau-Rayleigh instability in terms of the characteristic development time of the Plateau-Rayleigh instability and the corresponding wavelength, defined as:

$$\tilde{\tau} = -\frac{1}{\text{Im}(\tilde{\omega})_{max}} \quad \text{and} \quad \tilde{\lambda} = -\frac{2\pi}{\tilde{k}}. \qquad (12)$$

In Fig. 5, we show $\tilde{\tau}$ and corresponding $\tilde{\lambda}$ for $\tilde{\mu}$ in the range of 0-1.0 [5]. In dimension forms, the polynomial approximation of $\tilde{\tau}$ and $\tilde{\lambda}$ shown in Fig. 5 are:

$$\tau = \left(5.9982 \cdot \frac{\mu}{\sqrt{\gamma \cdot \rho \cdot R_0}} + 2.8778\right) \sqrt{\frac{\varrho R_0^3}{\gamma}}, \qquad (14)$$

$$\lambda = \left(0.1175 \cdot \left(\frac{\mu}{\sqrt{\gamma \cdot \rho \cdot R_0}}\right)^3 - 1.3197 \cdot \left(\frac{\mu}{\sqrt{\gamma \cdot \rho \cdot R_0}}\right)^2 + \frac{7.9471 \cdot \mu}{\sqrt{\gamma \cdot \rho \cdot R_0}} + 9.014\right) \cdot R_0. \tag{15}$$

As one can see from Eqs. (14) and (15), with an increase in $\mu$, $\tau$ and $\lambda$ increase; with an increase in $R_0$, $\tau$ increases; with an increase in $\gamma$, both $\tau$ and $\lambda$ decrease, and with an increase in $\rho$, $\tau$ increases and $\lambda$ decreases, which makes perfect physical sense.

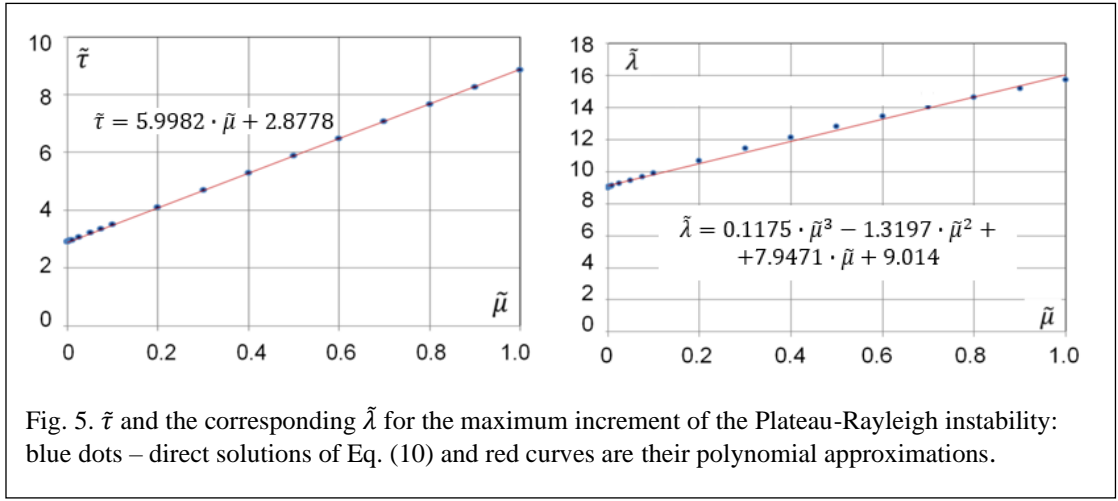

Fig. 5. $\tilde{\tau}$ and the corresponding $\tilde{\lambda}$ for the maximum increment of the Plateau-Rayleigh instability: blue dots – direct solutions of Eq. (10) and red curves are their polynomial approximations.

**D. Model of the ligament break-off**

In our model, the ligament ballistic regime (no Plateau-Rayleigh instability, Panel b in Fig. 1) switches to the Plateau-Rayleigh instability regime (Panel c in Fig. 1) where the following two conditions are fulfilled: (a) the length of the ligament is larger than wavelength of the instability $\lambda$ calculated at the current radius of the ligament, Eq. (15), and (b) the rate of decreasing the ligament radius due to the ballistic regime, $(dR/dt)_{Ball}$, is smaller than the characteristic rate of decreasing the radius of the bottleneck in the ligament due to the Plateau-Rayleigh instability, $(dR_{neck}/dt)_{P-R}$:

$$L \geq \lambda\left(\sqrt{\frac{V_{drop}}{\pi L}}\right), \tag{16}$$

$$\left|\left(\frac{dR}{dt}\right)_{Ball}\right| = \left|\frac{d}{dt}\left(\sqrt{\frac{V_{drop}}{\pi L}}\right)\right| = \sqrt{\frac{V_{drop}}{4\pi L^3}} \frac{dL}{dt} \leq \left|\left(\frac{dR_{neck}}{dt}\right)_{P-R}\right| = \frac{R}{\tau} = \frac{\sqrt{\frac{V_{drop}}{\pi L}}}{\tau\left(\sqrt{\frac{V_{drop}}{\pi L}}\right)}. \tag{17}$$

In Eqs. (16) and (17), we have substituted $R$, the radius of the ligament (Fig. 1), as $\sqrt{V_{drop}/\pi L}$, and, in Eq. (17) took $|(dR_{neck}/dt)_{P-R}|$ as $R/\tau$. Fig. 6 illustrates two scenarios of switching from the ballistic regime into the Plateau-Rayleigh instability regime.

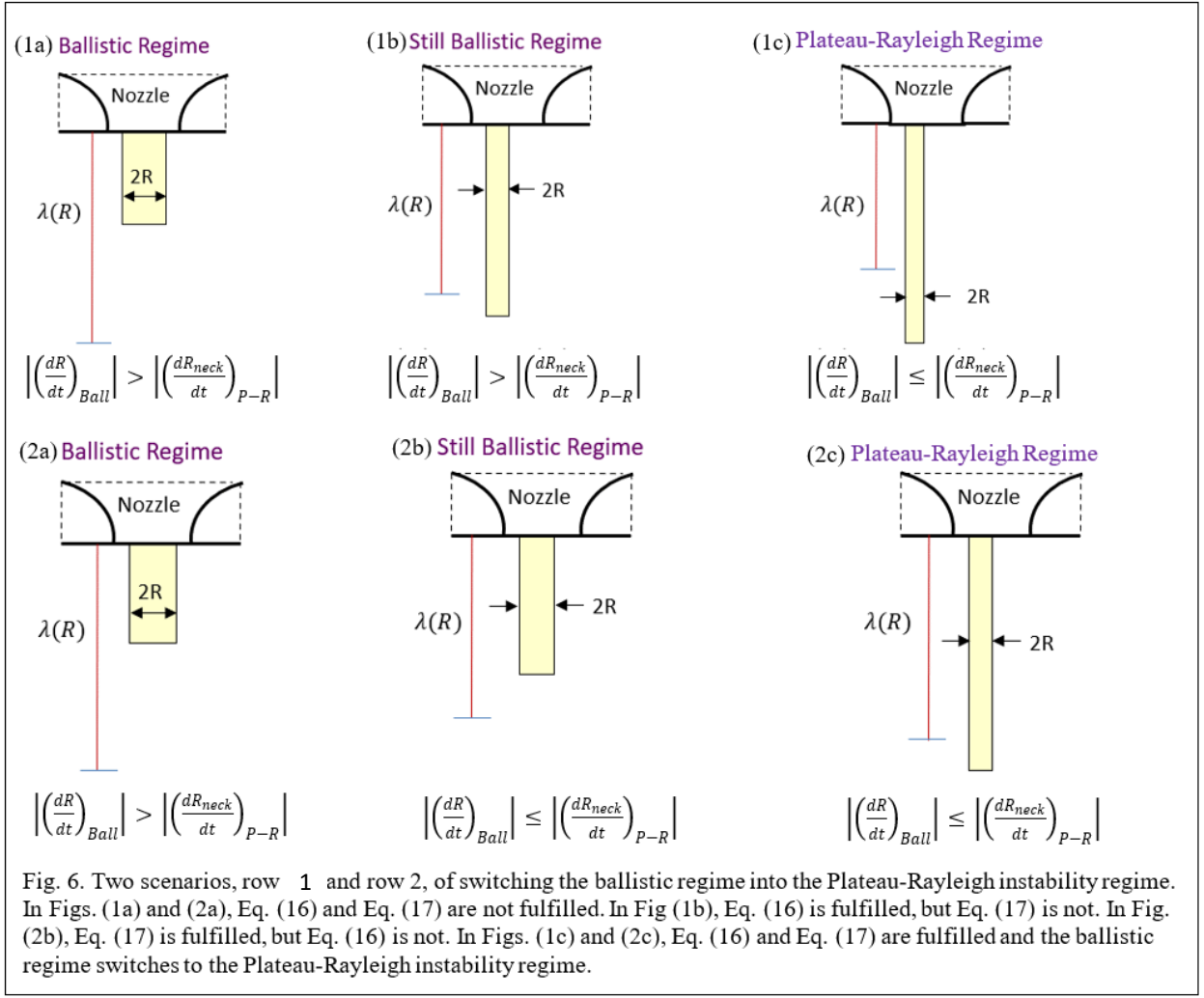

Fig. 6. Two scenarios, row 1 and row 2, of switching the ballistic regime into the Plateau-Rayleigh instability regime. In Figs. (1a) and (2a), Eq. (16) and Eq. (17) are not fulfilled. In Fig (1b), Eq. (16) is fulfilled, but Eq. (17) is not. In Fig. (2b), Eq. (17) is fulfilled, but Eq. (16) is not. In Figs. (1c) and (2c), Eq. (16) and Eq. (17) are fulfilled and the ballistic regime switches to the Plateau-Rayleigh instability regime.

In the model, we use the following ligament break-off algorithm:

(a) *Plateau-Rayleigh instability regime, functions vs. L:*

(a1) Calculate: $R(L) = \left(\frac{V_{drop}}{\pi \cdot L}\right)^{1/2}$.

(a2) Calculate: $\tilde{\mu}(L) = \frac{\mu}{\sqrt{\gamma \cdot \rho \cdot R(L)}}$.

(a3) Calculate: $\tau(L) = (5.9982 \cdot \tilde{\mu}(L) + 2.8778) \cdot \rho^{1/2} \cdot \gamma^{-1/2} \cdot (R(L))^{3/2}$.

(a4) Calculate: $\lambda(L) = (0.1175 \cdot \tilde{\mu}(L)^3 - 1.3197 \cdot \tilde{\mu}(L)^2 + 7.9471 \cdot \tilde{\mu}(L) + 9.014) R(L)$.

(a5) Calculate: $\left|\left(\frac{dR_{neck}}{dt}\right)_{P-R}\right| = \frac{R(L)}{\tau(L)}$.

(b) *Ballistic regime, functions vs time:*

(b1) Calculate $L(t)$ by using Eq. (8).

(b2) Calculate $\left|\left(\frac{dR}{dt}\right)_{Ball}\right| = \sqrt{\frac{V_{drop}}{4\pi L^3}} \frac{dL(t)}{dt}$.

(c) *Criteria for switching from the ballistic regime to the Plateau-Rayleigh instability regime:*

(c1) *if* $\{L(t) > \lambda(L(t))$ and $\left|\left(\frac{dR}{dt}\right)_{Ball}\right| < \left|\left(\frac{dR_{neck}}{dt}\right)_{P-R}\right|\}$ *then* $\{\tau_{Ball} = t;\ L_{Ball} = L;\}$ *else* $\{stop\}$.

(d) *Calculating the ligament break-off time, the ligament break-off length, and the velocity of the droplet* (Panels d and e in Fig. 1):

(d1) Calculate $\tau_{P-R} = (5.9982 \cdot \tilde{\mu}(L_{Ball}) + 2.8778) \cdot \rho^{1/2} \cdot \gamma^{-1/2} \cdot (R(L_{Ball}))^{3/2}$.

(d2) Calculate $t_{break-off} = \tau_{Ball} + \tau_{P-R}$.

(d3) Calculate $L_{break-off}$ by solving Eq. (8) until $t = t_{break-off}$.

(d4) Calculate $v_{drop} = \frac{1}{2}\left(\frac{dL(t)}{dt}\right)_{t_{break-off}}$ using Eq. (8).

In (d3) and (d4), we have assumed that the dynamic of the ligament lengthening in the Plateau-Rayleigh instability regime is the same as in the ballistic regime, assumption 8.

### III. Numerical Results, Comparison with Experiment

The time of squeezing liquid from the nozzle depends on the nozzle shape, the form of the pressure pulse, and other parameters. Therefore, a priori, we do not know the value of $\tau_{squeeze}$ which depends on the pressure pulse form, the nozzle shape, and other parameters. However, the duration of pressure pulse, $\tau_{pulse}$, is a reasonable design parameter for drop-on-demand printheads. To investigate the effect of $\tau_{squeeze}$ on the process of drop ejection, we calculate the droplet mass vs. drop velocity for a Dimatix SG1024-LCHF printhead for two cases: $\tau_{squeeze}/\tau_{pulse} = 1$ and 0.829 and compared the obtained results with the experimental data, Fig. 7. In this simulation, $R_{nozzle} = 25\mu m$, $\tau_{pulse} = 10\mu s$, the liquid is Prova@33C ($\rho = 950$ kg/m³, $\gamma = 0.028$ N/m, $\mu = 0.01$ Pa·s), and the ratio of $\tau_{squeeze}$ to $\tau_{pulse}$ is taken as 0.829 to match the calculated droplet velocity for $M_{drop} = 85.79$ng to its experiment value of $9.46 m/s$. The accuracy of the drop velocity measurement is about 0.2 m/s. As shown in Fig. 7, the differences in the droplet mass vs. drop velocity curves for $\tau_{squeeze}/\tau_{pulse} = 1$, and 0.829 are relatively small. For small $M_{drop}$, the code produces no solution. This was expected, because for such small droplet masses, the initial kinetic energy of the bulge is so small that the ligament rebounds back into the nozzle before the Plateau-Rayleigh instability is fully developed, Fig. 2. As one can see, the agreement between experiment data and the simulation is good in the whole region of measurements.

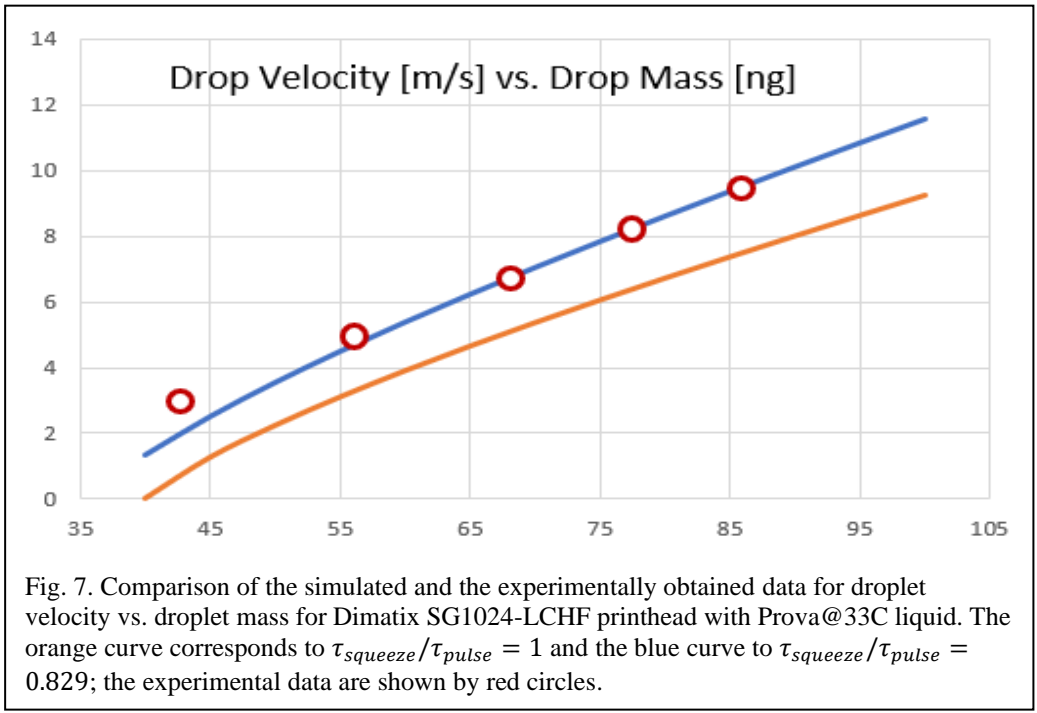

Fig. 7. Comparison of the simulated and the experimentally obtained data for droplet velocity vs. droplet mass for Dimatix SG1024-LCHF printhead with Prova@33C liquid. The orange curve corresponds to $\tau_{squeeze}/\tau_{pulse} = 1$ and the blue curve to $\tau_{squeeze}/\tau_{pulse} = 0.829$; the experimental data are shown by red circles.

The calculated ligament break-of length and the ligament break-off time vs. drop mass are shown in Fig. 8. Unfortunately, we do not have experimental data for ligament break-off length and break-off time to compare them against our numerical results. As one can see from Panel (b), for the same droplet masses, $t_{break-off}$, in the case of $\tau_{squeeze}/\tau_{pulse} = 0.829$, is smaller compared to the case of $\tau_{squeeze}/\tau_{pulse} = 1$. This was expected, because the ligament with larger initial face velocity, Eq. (4), compresses faster and, therefore, the Plateau-Rayleigh instability "detaches" the ligament from the nozzle earlier. At the same time, as follows from Panel (a), $L_{break-off}$ is longer for smaller $\tau_{squeeze}$ (for larger initial face velocity).

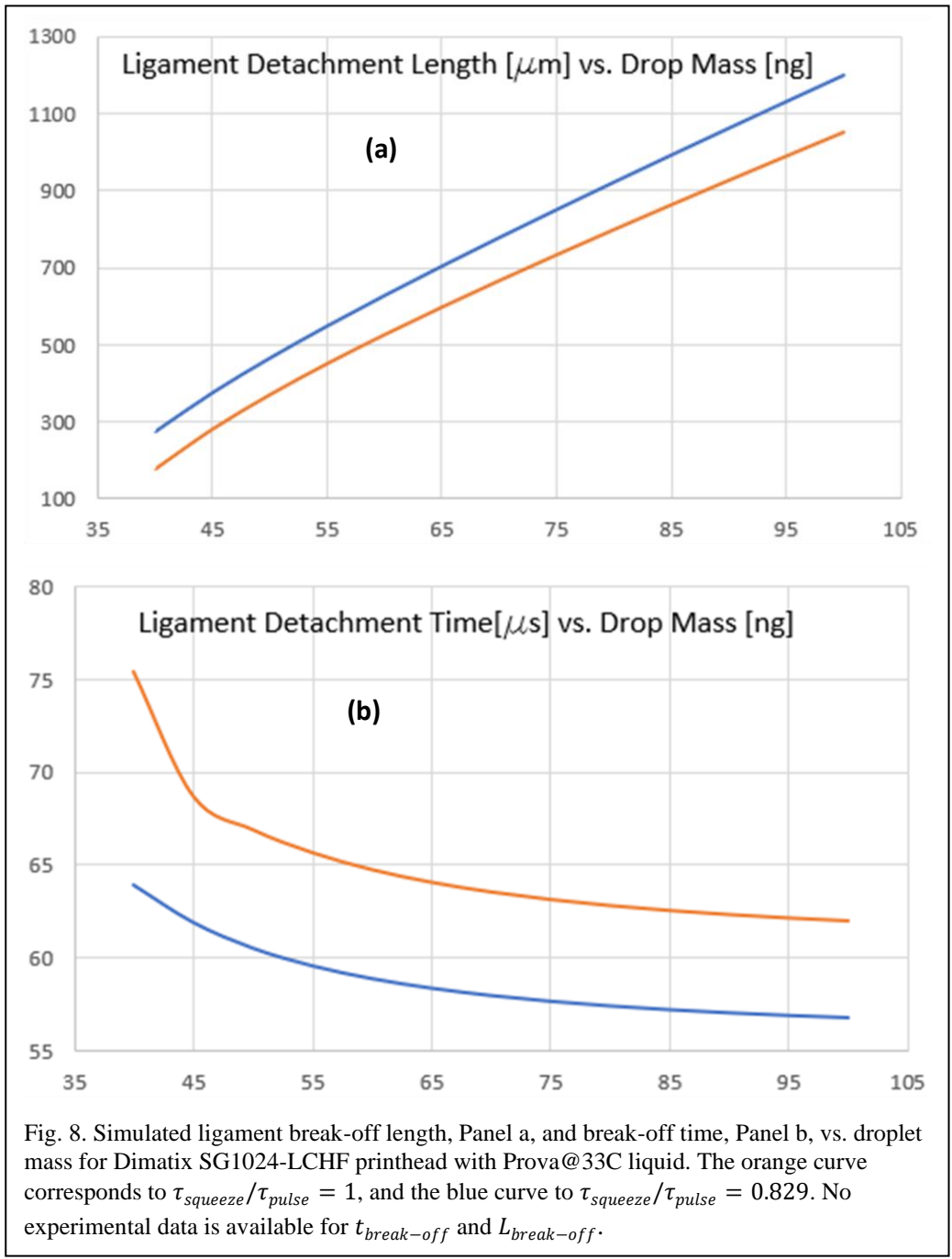

Fig. 8. Simulated ligament break-off length, Panel a, and break-off time, Panel b, vs. droplet mass for Dimatix SG1024-LCHF printhead with Prova@33C liquid. The orange curve corresponds to $\tau_{squeeze}/\tau_{pulse} = 1$, and the blue curve to $\tau_{squeeze}/\tau_{pulse} = 0.829$. No experimental data is available for $t_{break-off}$ and $L_{break-off}$.

Thus, we have demonstrated that our analytical model with $\tau_{squeeze} = \tau_{pulse}$ produces reasonable results and can be used as a first step for design of drop-on-demand printheads.

## IV. Conclusions

In this paper, we constructed a simple analytical one-dimensional model of drop ejection from a micro-size nozzle. In this model, the drop is ejected from the nozzle by applying a short pressure pulse that enables a droplet to be ejected from the nozzle. When pressure pulses are not present inside the jet, the liquid is held inside of the nozzle by surface tension. In the model, we assume that the characteristic time of squeezing liquid from the nozzle, $\tau_{squeeze}$, is about the time of the pressure pulse, that after that time, $t > \tau_{squeeze}$, no liquid exchange between the liquid bulge squeezed from the nozzle and liquid inside of the nozzle, and that the ligament is cylindrical. We demonstrated a great agreement of our model with experiments with optimized value of $\tau_{squeeze}$, and a reasonable agreement with $\tau_{squeeze} = \tau_{pulse}$. This model can be used as a first step for design of drop-on-demand printheads helping to understand the physical processes of drop formations in such systems.

## Acknowledgments

The author would like to express his sincere gratitude to his colleagues Dan Barnett, Sean Chahil, and Marlene MacDonald for their valuable comments, encouragement, and kind help in preparation the text of this article. The author thanks Paul Hoisington, James Myrick, Chris Menzel, James Cole-Henry, and Matthew Aubrey for helpful discussions.

## Appendix, Derivation of Eq. (5)

In derivation of Eq. (5), we will use the set of slender jet set equation [2] in form [3], Fig A1:

$$\frac{\partial v_z}{\partial t} + v_z \frac{\partial v_z}{\partial z} = -\frac{\gamma}{\rho}\frac{\partial H}{\partial z} + \frac{3\mu}{\rho}\frac{1}{y}\frac{\partial}{\partial z}\left(y\frac{\partial v}{\partial z}\right), \tag{A1}$$

$$\frac{\partial}{\partial t}(y) + \frac{\partial}{\partial z}(y v_z) = 0, \tag{A2}$$

$$y = R^2, \tag{A3}$$

$$H = \frac{1}{R_1} + \frac{1}{R_2} = \frac{y - \frac{y}{2}\frac{\partial^2 y}{\partial z^2} + \frac{1}{2}\left(\frac{\partial y}{\partial z}\right)^2}{\left(y + \frac{1}{4}\left(\frac{\partial y}{\partial z}\right)^2\right)^{3/2}}, \tag{A4}$$

where $R(z,t)$ and $v_z(z,t)$ are the radius and the flow velocity of the jet respectively. Eq. (A1) is the jet z-momentum equation, Eq. (2) is the volume conservation equation, and Eq. (A4) describe the curvature at the surface of the jet. The first term in RHS of Eq. (A1) describes the surface tension forces and the second one the viscosity forces. Boundary conditions for Eqs. (A1) and (A2) are:

$$v_z(0,t) = 0 \quad \text{and} \quad y(0,t) = y_0(t), \tag{A5}$$

$$v_z(L,t) = \frac{dL}{dt} \quad \text{and} \quad y(L,t) = 0. \tag{A6}$$

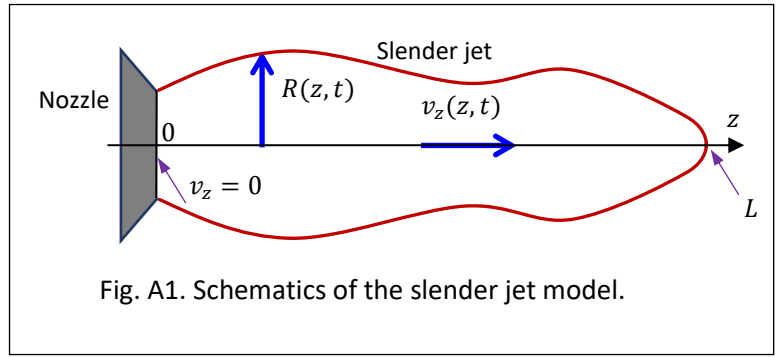

Fig. A1. Schematics of the slender jet model.

As one can see they correspond to boundary conditions that we have used in our analytical "cylindrical" ligament model, Fig. 1, and Eqs. (2) and (3).

Multiplying Eq. (A1) by $\pi \rho v y$ and then integrating the obtained equation over the length of the jet, we obtain an energy conservation equation for jet:

$$\pi \rho v_z y \left(\frac{\partial v_z}{\partial t} + v \frac{\partial v_z}{\partial z}\right) = -\pi v y \gamma \frac{\partial H}{\partial z} + 3\pi \mu v \frac{\partial}{\partial z}\left(y \frac{\partial v}{\partial z}\right) \underset{v_z = 0}{\rightarrow}$$

$$\rightarrow \pi \frac{\partial}{\partial t}\left(\frac{\rho v_z^2 y}{2}\right) + \pi \rho \frac{\partial}{\partial z}\left(\frac{v_z^3 y}{2}\right) = -\pi \gamma \frac{\partial}{\partial z}(v_z y H) + \pi \gamma H \frac{\partial}{\partial z}(v_z y) + 3\pi \mu v_z \frac{\partial}{\partial z}\left(y v_z \frac{\partial y}{\partial z}\right) - 3\pi \mu y \left(\frac{\partial v_z}{\partial z}\right)^2 \rightarrow$$

$$\rightarrow \pi \frac{\partial}{\partial t}\left(\frac{\rho v_z^2 y}{2}\right) + \pi \rho \frac{\partial}{\partial z}\left(\frac{v_z^3 y}{2}\right) = -\pi \gamma \frac{\partial}{\partial z}(v_z y H) - \pi \gamma H \frac{\partial y}{\partial t} + 3\pi \mu v_z \frac{\partial}{\partial z}\left(y v_z \frac{\partial y}{\partial z}\right) - 3\pi \mu y \left(\frac{\partial v_z}{\partial z}\right)^2 \tag{A7}$$

In derivation of Eq. (A7), we have used Eq. (A2). The integration Eq. (A7) over the length of the jet yields:

$$\pi \int_0^L \frac{\partial}{\partial t}\left(\frac{\rho v_z^2 y}{2}\right) dz + \pi \left(\frac{\rho v_z^3 y}{2}\right)\bigg|_0^L = -\pi \gamma \{v_z y H\}_0^L - \pi \gamma \int_0^L H \frac{\partial y}{\partial t} dz + \left\{3\pi \mu v_z h \frac{\partial y}{\partial z}\right\}_0^L - 3\mu \pi \int_0^L \left(\frac{\partial v_z}{\partial z}\right)^2 y\, dz \rightarrow$$

$$\rightarrow \pi \int_0^L \frac{\partial}{\partial t}\left(\frac{\rho v_z^2 y}{2}\right) dz = -\pi\gamma \int_0^L H \frac{\partial y}{\partial t} dz - 3\mu\pi \int_0^L \left(\frac{\partial v_z}{\partial z}\right)^2 y dz \rightarrow$$

$$\rightarrow \pi \frac{\partial}{\partial t}\left(\int_0^L \frac{\rho v_z^2 y}{2} dz\right) - \pi \left(\frac{\rho v_z^2 y}{2}\right)_L \frac{\partial L}{\partial t} = -\pi\gamma \int_0^L H \frac{\partial y}{\partial t} dz - 3\mu\pi \int_0^L \left(\frac{\partial v_z}{\partial z}\right)^2 y dz \rightarrow$$

$$\rightarrow \frac{\partial}{\partial t}\left(\int_0^L \frac{\pi \rho v_z^2 y}{2} dz\right) = -\pi\gamma \int_0^L H \frac{\partial y}{\partial t} dz - 3\mu\pi \int_0^L \left(\frac{\partial v_z}{\partial z}\right)^2 y dz \quad (A8)$$

In derivation of Eqs. (A8), we have taken into account BC (A5) and (A6). Eq. (8) describes the kinetic energy losses due to work against the surface tension and viscosity forces.

Substitution $H = \frac{1}{R}$, $y = R^2$, and $v_z = \frac{z}{L}\frac{dL}{dt}$ into Eq. (A9) yields the following equation describing the ligament dynamics:

$$\frac{d}{dt}\left(\frac{\pi\rho}{2} R^2 \left(\frac{1}{L}\frac{dL}{dt}\right)^2 \int_0^L z^2 dz\right) = -2\pi\gamma \int_0^L \frac{dR}{dt} dz - \frac{3\mu\pi R^2}{L^2}\left(\frac{dL}{dt}\right)^2 \int_0^L dz \rightarrow$$

$$\rightarrow \frac{d}{dt}\left(\frac{\pi\rho}{6} R^2 L \left(\frac{dL}{dt}\right)^2\right) = -2\pi\gamma \frac{d}{dt}\left(\int_0^L R dz\right) + 2\pi\gamma R(z=L)\frac{dL}{dt} - \frac{3\mu\pi R^2}{L}\left(\frac{dL}{dt}\right)^2 \rightarrow$$

$$\rightarrow \frac{d}{dt}\left(\frac{\pi\rho}{6} R^2 L \left(\frac{dL}{dt}\right)^2\right) = -2\pi\gamma \frac{d}{dt}(RL) - \frac{3\mu\pi R^2}{L}\left(\frac{dL}{dt}\right)^2 \rightarrow$$

$$\rightarrow \frac{d}{dt}\left(\frac{\pi\rho}{6} R^2 L \left(\frac{dL}{dt}\right)^2 + 2\pi\gamma RL\right) = -\frac{3\mu\pi R^2}{L}\left(\frac{dL}{dt}\right)^2 \quad (A9)$$

In derivation of Eq. (A9), we have taken into account that $R(z = L)$ is zero, Fig. A1, and neglected the area of the "flat" meniscus at the tip of the ligament, Fig. 1. Substituting $R = \sqrt{V_{ligament}/\pi L}$ into Eq. (A9) yields Eq. (5).